\documentclass[apj]{emulateapj}
\usepackage[colorlinks,linkcolor={blue},citecolor={blue},urlcolor={red}]{hyperref}
 \usepackage{epsfig} \usepackage{graphicx}
\usepackage{natbib} \usepackage{amsmath} \usepackage{amsfonts}
\usepackage{amssymb}

\newcommand{\myemail}{\email{leech@shao.ac.cn}}

\slugcomment{To appear in ApJ Letters}
\shorttitle{Large scale alignment of massive galaxies at $z\sim0.6$}
\shortauthors{C. Li et al.}

\begin{document}

\title{Detection of the large scale alignment of massive galaxies at
  $z\sim0.6$}

\author{Cheng  Li\altaffilmark{1}, Y. P.  Jing\altaffilmark{2},
  A. Faltenbacher\altaffilmark{3}, and Jie Wang\altaffilmark{4}}
\myemail

\altaffiltext{1}{Partner   Group   of  the   Max   Planck  Institute
  for Astrophysics  at  the  Shanghai  Astronomical  Observatory  and
  Key Laboratory for Research in Galaxies and Cosmology of Chinese
  Academy of   Sciences,    Nandan   Road   80,    Shanghai   200030,
  China} \altaffiltext{2}{Center for Astronomy and Astrophysics,
  Department of Physics, Shanghai Jiao Tong University, Shanghai
  200240, China} \altaffiltext{3}{School of Physics, University of the
  Witwatersrand, PO Box Wits, Johannesburg 2050, South Africa}
\altaffiltext{4}{National  Astronomical Observatories,  Chinese
  Academy of  Sciences, Beijing  100012, China}

\begin{abstract}
We report on the detection of the alignment between galaxies and
large-scale structure at $z\sim0.6$ based on the CMASS galaxy sample
from the Baryon Oscillation Spectroscopy Survey data release 9.  We
use two statistics to quantify the alignment signal: 1) the alignment
two-point correlation function which probes the dependence of  galaxy
clustering at a given separation in redshift space on  the projected
angle ($\theta_p$) between the orientation of galaxies and the line
connecting to other galaxies, and 2) the $\cos(2\theta)-$statistic
which estimates the average of $\cos(2\theta_p)$ for all correlated
pairs at given separation $s$.  We find significant alignment signal
out to about 70 $h^{-1}$Mpc in both statistics.  Applications of the
same statistics to dark matter halos of mass above $10^{12} h^{-1}M_\odot$ 
in a large cosmological simulation  show similar scale-dependent alignment
signals to the observation,  but with higher amplitudes at all scales
probed. We show that this discrepancy may be partially explained by a
misalignment angle between central galaxies and their host halos,
though detailed modeling is needed in order to better understand the
link between the orientations of galaxies and host halos. In addition,
we find systematic trends of the alignment statistics with the stellar
mass of the CMASS galaxies, in the sense that more massive galaxies 
are more strongly aligned with the large-scale structure.
\end{abstract}

\section{Introduction}
\label{sec:introduction}

Galaxies are not oriented at random, but show various forms of spatial
alignment \citep{Carter-Metcalfe-80, Binggeli-82, Dekel-85,  West-89b,
  Struble-90, Plionis-94, Plionis-03,
  Hashimoto-Henry-Boehringer-08}. In particular, recent studies of
galaxies in the Sloan Digital Sky Survey \citep[SDSS;][]{York-00} have
revealed that satellite galaxies are preferentially distributed along
the major axis of the central galaxies \citep{Brainerd-05, Yang-06a,
  Azzaro-07, Faltenbacher-07, Faltenbacher-09}, and tend to be
preferentially oriented toward the central galaxy
\citep{Pereira-Kuhn-05, Agustsson-Brainerd-06a,
  Faltenbacher-07}. These studies were mostly  limited to the local
universe and intermediate-to-small scales (less than a few tens of Mpc).
Similar alignment signals have been detected for
galaxies  at intermediate redshifts
\citep[$0.2<z<0.5$;][]{Donoso-O'Mill-Lambas-06, Okumura-Jing-Li-09}.
In a recent study \citet{Smargon-12} reported on the detection of
intrinsic alignment between clusters of galaxies at $0.08<z<0.44$ out
to 100 $h^{-1}$Mpc in the SDSS cluster catalogs. 
There have also been many observational studies which include galaxy
ellipticity and measure both the galaxy orientation-density correlation
and the intrinsic shear-density correlation \citep[e.g.][]{Mandelbaum-06a, 
Hirata-07, Blazek-McQuinn-Seljak-11, Joachimi-11}, in order to better
understand the potential contamination of galaxy alignment to cosmic 
shear surveys.

The various forms of the alignment of galaxies and clusters are
generally expected in the $\Lambda$ cold dark matter cosmological
paradigm in which galaxies are hosted by dark matter halos and are
embedded in a cosmic web containing  a varity of structures.  The
shapes of halos (and galaxies) are predicted to be aligned  with each
other due to the large-scale tidal field and the preferred accretion
of matter along filaments \citep{Pen-Lee-Seljak-00, Croft-Metzler-00,
  Heavens-Refregier-Heymans-00, Catelan-Kamionkowski-Blandford-01,
  Crittenden-01, Jing-02, Porciani-Dekel-Hoffman-02b}. Therefore,
measuring the alignment of galaxies/clusters, as function of redshift,
spatial scale and galaxy properties, is expected to provide useful
constraints on both galaxy formation and structure formation models.

In this work we extend the effort of detecting galaxy alignment 
to higher redshifts ($0.4<z<0.7$) and larger scales
($<200 h^{-1}$Mpc). For this we use the recently-released CMASS galaxy
sample from the ninth data release \citep[DR9;][]{Ahn-12} of the
Baryon  Oscillation Spectroscopic Survey
\citep[BOSS;][]{Schlegel-White-Eisenstein-09, Dawson-13}, which is  a
part of the SDSS-III \citep{Eisenstein-11}.  We apply two different
statistics suitable for quantifying the spatial alignment of galaxies
to the CMASS sample, and show that  the alignment between the
orientation of the CMASS galaxies and the large-scale galaxy
distribution extends out to $120 h^{-1}$Mpc. Applying the same
statistics to dark matter halos in a large cosmological  simulation,
we detect similar alignment signals for the halos. This indicates that
the observed large-scale alignment of galaxies can be explained by the
anisotropy in the large-scale matter distribution, as we have recently
found from a theoretical analysis of dark matter halos
\citep{Faltenbacher-Li-Wang-12}.

\section{Methodology}

We use two different statistics to quantify the alignment between the
orientation of galaxies and their large-scale spatial distribution:
the alignment correlation function (ACF) and the
$\cos(2\theta)-$statistic, which were originally introduced in
\citet{Faltenbacher-09}. Here we briefly describe the statistics and
refer the reader to that paper for details.

\subsection{Alignment correlation function}

The ACF extends the conventional two-point correlation function 
(2PCF) by including the angle between the major axis of a galaxy
and the line connecting to another galaxy ($\theta_p$, projected 
on the sky for a survey sample) as an 
additional property of galaxy pairs. For a pair of galaxies with 
one member in the sample in question (called Sample Q hereafter) and 
another member in the reference sample (called Sample G hereafter), 
we consider $\theta_p$ as a secondary property of the pair, in 
addition to the separation of the paired galaxies. The estimator
for the conventional 2PCF is then easily modified to give a
measure of the ACF:
\begin{equation}\label{eqn:estimator}
  \xi(\theta_p,s) = \frac{N_{R}}{N_{G}}
  \frac{QG(\theta_p,s)}{QR(\theta_p,s)} -1,
\end{equation}
where $s$ is the redshift-space pair separation, $N_G$ and
$N_R$ are the number of galaxies in the reference and random
samples. $QG(\theta_p,s)$ and $QR(\theta_p,s)$ are the counts of cross
pairs between the given samples for given $\theta_p$ and $s$. The
value of $\theta_p$ ranges from zero (parallel to the major axis
of the main galaxy) to 90 degrees (perpenticular). Thus, higher
amplitudes of $\xi(\theta_p,s)$ at small (large) $\theta_p$ indicate
the galaxies in G are preferentially aligned along the
major (minor) axis of the galaxies in Q. Sample Q is either the same
as, or a subset of Sample G. In the former case the ACF is actually
the alignment {\em auto}-correlation function,  thus probing the
alignment between galaxies within the same sample.

\subsection{The $\cos(2\theta)-$statistic}

The $\cos(2\theta)-$statistic measures the average value of $\cos(2\theta)$
over all {\em correlated} pairs for a given spatial separation. 
This statistic is related  to the ACF by
\begin{equation}
  \langle\cos(2\theta_p)_{\mbox{cor}}\rangle (s) =
  \frac{\int_0^{\pi/2}\cos(2\theta_p)\xi(\theta_p,s)d\theta_p}
       {\int_0^{\pi/2}\xi(\theta_p,s)d\theta_p},
\end{equation}
and estimated by
\begin{equation}
  \langle \cos(2\theta_p)_{\mbox{cor}} \rangle(s) =
  \frac{QG_{\theta_p}(s)}{QG(s)-(N_G/N_R)\cdot QR(s)},
\end{equation}
where $QG_{\theta_p}(s)$ is the sum of $\cos(2\theta_p)$ for all  the
cross pairs between samples  $Q$ and $G$ at separation $s$:
\begin{equation}
  QG_{\theta_p}(s) = \sum_{(i,j)\in QG(s)}\cos(2\theta_{p}^{i,j}).
\end{equation}
The statistic so-defined ranges between -1 and
1, with positive and negative values indicating a preference for small
($<45^\circ$) and large ($>45^\circ$) angles. Values of zero means isotropy. 

\section{Data}

\subsection{The BOSS/CMASS galaxy sample}

By selection the CMASS is a roughly volume-limited sample of massive
galaxies in the redshift range of $0.4<z<0.7$  \citep{Eisenstein-11,
  Anderson-12}.  Clustering measurements and halo occupation
distribution modeling \citep{White-11} revealed that the CMASS
galaxies are hosted  by dark matter halos with mass above
$10^{12}h^{-1}M_\odot$, with  the majority ($90\%$) being central
galaxies in halos of mass $\sim10^{13}h^{-1}M_\odot$. The CMASS galaxy
sample from BOSS/DR9 and the corresponding random sample suitable for
large-scale structure analyses are generated by \citet{Anderson-12},
and are publicly available  at the SDSS-III
website\footnote{http://data.sdss3.org/datamodel/fiels/BOSS\_LSS\_REDUX/}.
For this work we restrict ourselves to the survey area in the northern
Galactic cap, including a total number of 207,246 galaxies. We exclude
the sourthern Galactic cap from our analysis in order to avoid
possible effects of the systematic differences  between the northern
and southern parts as found in recent BOSS-based studies
\citep[e.g.][]{Sanchez-12a}.  The orientation of the galaxies is
given by the position angle (PA) of the major axis of their
$r$-band images, determined from the de Vaucouleurs model fit
by the SDSS photometric pipeline {\sc Photo}\citep{Lupton-01, Stoughton-02}.

\subsection{The MultiDark Run 1 Simulation}

In addition to analyzing the CMASS galaxy sample, we also apply our
alignment statistics to dark matter halos in the MultiDark Run 1
simulation \citep[MDR1;][]{Prada-12} 
\footnote{http://www.multidark.org/MultiDark/}.  Assuming the WMAP7
concordant $\Lambda$CDM cosmology, the simulation uses $2048^3$
particles to follow the dark matter distribution in a cubic region
with 1 $h^{-1}$Gpc on a side, which corresponds to a particle mass of
$8.72\times10^9h^{-1}M_\odot$. Dark matter halos are identified by
means of a friends-of-friends \citep{Davis-85} algorithm with a
linking length of 0.17 times the mean particle separation.  For the
comparison with the CMASS galaxy sample  we use snapshot 60 which
corresponds to a redshift of $z\sim0.6$. 

Following \citet{Joachimi-13} we determine the projected
orientations of the dark matter halos based on the 3D mass ellipsoids
which are provided in the Multidark database. For this approach the
line of sight is assumed to be parallel to the $z$-axis.
We limit our analysis to  dark matter
halos with masses above   $10^{12}h^{-1}M_\odot$, i.e., the
aforementioned lower limit of the host  halo mass for CMASS galaxies
as found by \citet{White-11}.  Halos of this  mass are identified with
a number of 115 particles. In this case an uncertainty of 10\% is
expected for the halo orientation determination \citep{Bett-07, Joachimi-13},
which we expect not to introduce significant bias into our result
in the next section.

\section{Results}

\begin{figure*}
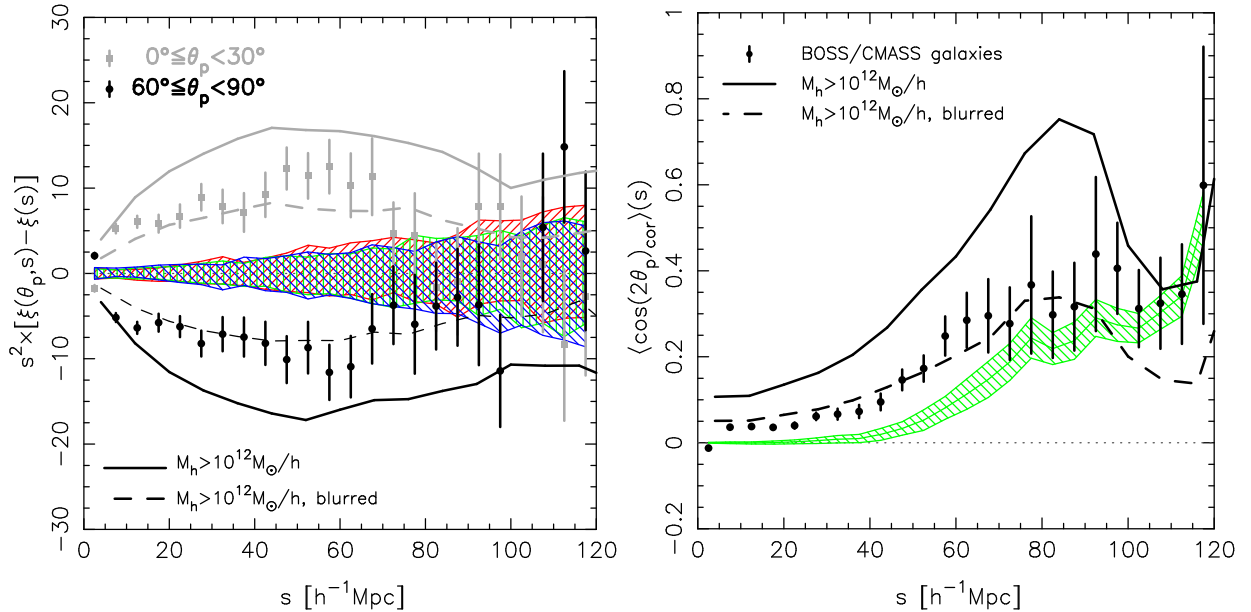

  \begin{center}
    \epsfig{figure=f1a.ps,width=0.45\hsize}
    \epsfig{figure=f1b.ps,width=0.45\hsize}
  \end{center}
  \caption{{\it Left:} difference between the alignment correlation
    function at given projected angle $\xi(\theta_p,s)$ and the
    conventional correlation function $\xi(s)$, obtained from the
    CMASS galaxy sample. Results at the small and large $\theta_p$
    bins are plotted in grey and black symbols separately. The
    hatched regions plotted in red/green/blue represent the $1\sigma$
    variance between 100 random samples in which the position angles
    are randomly shuffled among the galaxies, measured for the three
    angle bins separately. The solid
    and dashed lines show the results for dark matter halos with mass
    above $10^{12}h^{-1}M_\odot$. The solid lines are for the halos
    with no misalignment, and the dashed lines  are results with the
    misalignment parameter of $\sigma_\theta=35^\circ$.  {\it Right:}
    the $\cos(2\theta)-$statistic measured for the same galaxy sample
    and dark matter halo catalog. Symbols and lines are the same as in
    the left-hand panel.}
  \label{fig:ratio_xis_all}
\end{figure*}

We have obtained the alignment correlation function $\xi(\theta_p,s)$ 
from the CMASS galaxy sample for three successive angular intervals: 
$0^\circ\leq\theta_p<30^\circ$, $30^\circ\leq\theta_p<60^\circ$, and 
$60^\circ\leq\theta_p<90^\circ$, as well as the conventional two-point 
correlation function, $\xi(s)$, which is a function of only the 
redshift-space separation and can be regarded as an average of the 
alignment correlation function over the full range of $\theta_p$.  

In Figure~\ref{fig:ratio_xis_all} (left panel) we plot the
difference in the alignment correlation function at small/large angles
with respect to the conventional correlation function $\xi(s)$.
The error bars plotted in the figure and in what follows are estimated
using the bootstrap resampling technique \citep{Barrow-Bhavsar-Sonoda-84}.
We have constructed 100 bootstrap samples based on the real sample,
and we estimate the difference between $\xi(\theta,s)$ and $\xi(s)$
for each sample. The error at given scale is then estimated from the
$1\sigma$ variance between the bootstrap samples.

As can be seen, $\xi(\theta_p,s)$ differ from $\xi(s)$ at both small and
large angles, with stronger clustering at smaller angles and weaker
clustering at larger angles, consistent with the picture that the 
major axis of the galaxies is preferentially aligned with their spatial
distribution. 

It is essential to perform systematics tests on any clustering measurements
\citep[e.g.][]{Mandelbaum-05, Sanchez-12b}. As one of such tests, we 
have repeated the
same analysis as above for a set of 100 random samples in which the 
position angles are shuffled at random among the main galaxies (Sample Q). 
The hatched regions plotted in red/green/blue 
in Figure~\ref{fig:ratio_xis_all} show the $1\sigma$ variance of the
alignment correlation function between the random samples, measured
for the three angle bins separately. It is interesting that the alignment signal
detected in the real sample is significantly seen for a wide range of scales,
from the smallest scales probed ($\sim 5 h^{-1}$Mpc) out to $\sim70 h^{-1}$Mpc
according to both the bootstrap errors of the measurements and the 
$1\sigma$ regions of the random samples.

The right-hand panel of Figure~\ref{fig:ratio_xis_all} shows the
$\cos(2\theta)-$statistic, plotted in solid circles for the CMASS
sample and in green hatched region for the 100 randomly shuffled samples.
The statistic for the real sample shows positive values on all scales probed, 
while its difference from the random samples is significantly seen only
for scales below $\sim70 h^{-1}$Mpc, consistent with what the left-hand
panel reveals. As mentioned above, a positive value in the 
$\cos(2\theta)-$statistic indicates a preference for angles smaller
than $45^\circ$, thus implying that the major axis of the galaxies
tends to be aligned with the large-scale distribution of galaxies.
The $\cos(2\theta)-$statistic of the random samples shows a systematic
positive bias at scales above $\sim40 h^{-1}$Mpc, implying that 
the position angle of the CMASS galaxies is not randomly distributed
on the sky, a cosmic variance effect due to the limited survey area
and probably also the quite irregular shape of the survey geometry.
This can be tested in future with mock catalogs or later data releases
of the BOSS survey.

\begin{figure*}
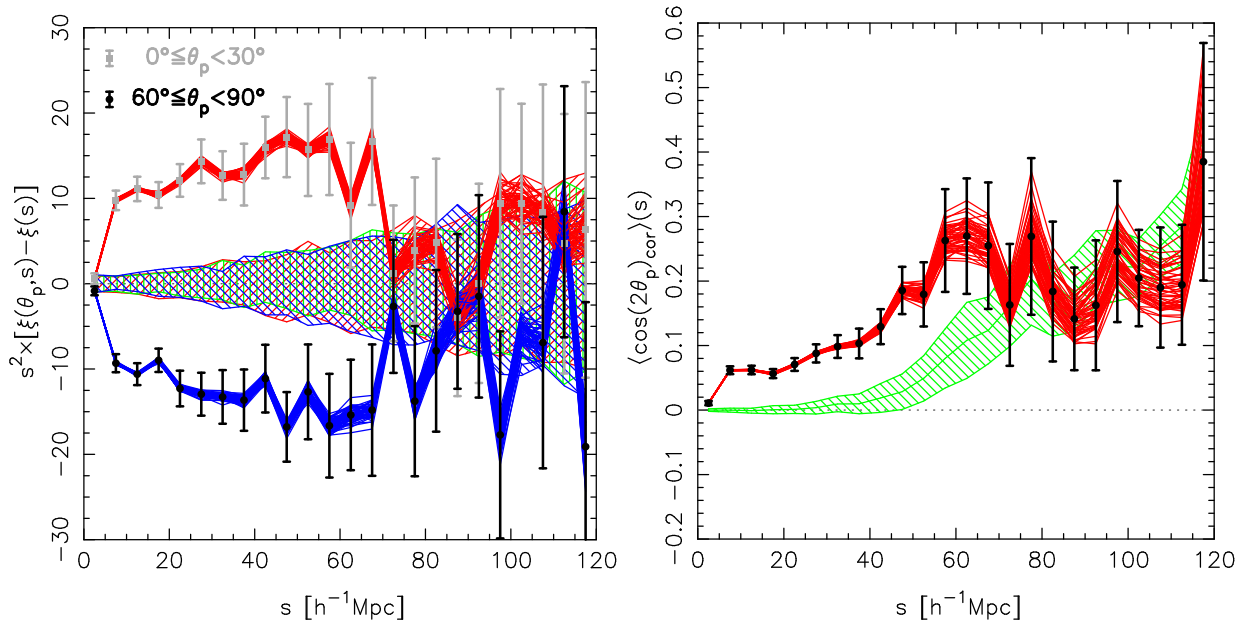

  \begin{center}
    \epsfig{figure=f2a.ps,width=0.45\hsize}
    \epsfig{figure=f2b.ps,width=0.45\hsize}
  \end{center}
  \caption{Alignment correlation function and the
    $\cos(2\theta)-$statistic for a subset of the CMASS galaxies 
    with {\tt fracDev}$>0.8$, $R_{deV}>1^{\prime\prime}$ and
    ellipticity $1-b/a>0.2$. Symbols and lines are the same 
    as in Figure 1, except that the red/blue lines in the left
    panel and the red lines in the right panel show the results
    for a set of 51 jacknife samples. See the text for detailed
    description.
    }
  \label{fig:sys_tests}
\end{figure*}

\begin{figure*}
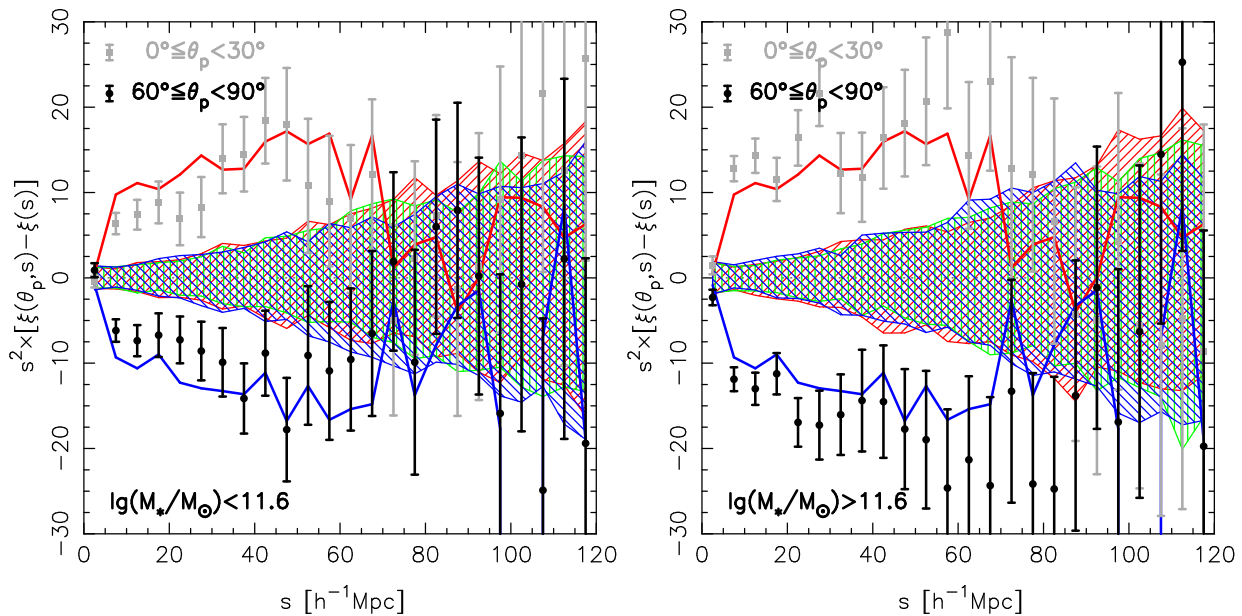

  \begin{center}
    \epsfig{figure=f3a.ps,width=0.45\hsize}
    \epsfig{figure=f3b.ps,width=0.45\hsize}
  \end{center}
  \caption{Alignment correlation function for the CMASS galaxies 
    with stellar masses either below (left panel) or above (right panel)
    $10^{11.6}M_\odot$. The result for the full sample is repeated
    for reference in both panels as the red/blue solid lines.}
  \label{fig:mass_dependence}
\end{figure*}

For comparison the same statistics obtained for dark matter halos of
mass $M_h>10^{12}h^{-1}M_\odot$ in the MDR1 simulation are shown in
Figure~\ref{fig:ratio_xis_all} as solid lines. Alignment signal is
seen in both statistics and on all the scales up to 120 $h^{-1}$Mpc.
Both statistics show strong dependence on the spatial scale, which is
very similar to what is seen for the CMASS galaxies. At fixed scale,
however, the alignment of the halos is systematically stronger than
that of the galaxies.  This discrepancy might be partially (if not
totally) due to  the misalignment between the orientation of central
galaxies and that of their host halos.  A previous study done by
\citet{Okumura-Jing-Li-09} on the alignment of luminous red galaxies
(LRGs) at $0.16<z<0.47$ in the SDSS/DR6 suggested that the
misalignment angle between a central LRG and its host halo follows a
Gaussian distribution with a zero mean and a typical width
$\sigma_\theta=35.4$ deg \citep[see also][]{Faltenbacher-09}.  Such
misalignment is expected to smooth out the alignment to some extent,
leading the alignment of galaxies to be weaker than that of their host
halos. Those authors found no evidence for the redshift evolution of
their results. Thus, it is likely that a similar misalignment occurs
also at the redshift of the CMASS sample. 

To test the effect of such misalignment on our statistics, we
artificially assign misalignment to the orientation of the dark matter
halo before the alignment statistics are measured, assuming a Gaussian
distribution with $\sigma_\theta=35^\circ$ for the misalignment angle.
The results are plotted as dashed lines in
Figure~\ref{fig:ratio_xis_all}.  As expected, the amplitude of both
statistics decreases considerably, becoming comparable with the data
on scales below $\sim70 h^{-1}$Mpc.  This simple experiment
implies that the observed large-scale alignment signal for the massive
galaxies at $z\sim0.6$ is real and can be explained by the alignment
between dark matter halos and the large-scale matter distribution
after the misalignment between the galaxies and their halos has been
considered properly.

In order to test whether the uncertainties in the position angle
measurements of the CMASS galaxies can introduce any systematic errors,
we have repeated the analysis for a subset of $\sim73,000$ galaxies
with the weight of the de Vaucouleurs model component {\tt fracDev}$>0.8$,
the de Vaucouleurs model scale radius $R_{deV}>1^{\prime\prime}$, 
and ellipticity $1-b/a>0.2$. The results are shown in 
Figure~\ref{fig:sys_tests}, with the grey/black symbols for the
real sample and the hatched regions for the randomly shuffled samples.
In addition, we have constructed 51 jacknife samples by dividing
the CMASS/North area into 51 non-overlapping subregions and dropping
one of the subregions from each of the 51 samples. The results
of these samples are plotted in the red/blue lines in the same figure.
These tests demonstrate that the alignment signal is 
reliably detected in the CMASS sample, at least to $\sim70 h^{-1}$Mpc,
and is robust to the position angle measurements and the statistical
error estimation.

Finally, we focus on the CMASS galaxies and examine the dependence of
the alignment statistics on the stellar mass of the galaxies. For this
we take the stellar mass esimates from the Wisconsin group
\footnote{http://www.sdss3.org/dr9/algorithms/galaxy\_wisconsin.php}
derived by \citet{Chen-12} from a BOSS spectrum principal component
analysis (PCA) using the stellar population models of
\citet{Bruzual-Charlot-03}. We divide the CMASS galaxies into two
subsamples, with stellar mass either below or above $10^{11.6}M_\odot$.
We take each of the subsamples as Sample Q, and we measure the alignment
cross-correlation function with respect to the full CMASS galaxy sample
(Sample G) in the same way as above. The results are shown in 
Figure~\ref{fig:mass_dependence}, with the two panels for the two
subsamples separately. The result of the full sample is repeated in
both panels as red/blue solid lines, for reference. Both subsamples
show systematic differences from the full sample, in the sense that
the high-mass subsample shows stronger-than-average alignment signals
and the low-mass subsample shows weaker-than-average signals. 

\section{Summary}

We have applied two statistics, that are defined to be suitable for
quantifying the spatial alignment of galaxies, to the CMASS galaxy
sample from the SDSS-III/BOSS DR9, which consist of  about
$2\times10^{5}$ massive galaxies with mass above  $\sim10^{11}M_\odot$
and redshift in the range $0.4<z<0.7$. Both statistics have revealed
significant alignment, out to $\sim70 h^{-1}$Mpc, between the major 
axis of the CMASS galaxies and the large-scale distribution of the 
galaxies in the same sample. In addition, we have also detected a 
systematic trend of the alignment with the stellar mass of the galaxies, 
in the sense that more massive galaxies are more strongly aligned with
the large-scale structure.

We have applied the same statistics to dark matter halos with mass
above $10^{12}h^{-1}M_\odot$ in the MultiDark Run 1 (MDR1) simulation,
and obtained very similar alignment sginals to what we have seen  for
the CMASS galaxies. This is consistent with previous studies of  halo
occupation distribution models on the CMASS sample which  indicated
that the majority of the CMASS galaxies are central galaxies in halos
of mass $M_h>10^{12}h^{-1}M_\odot$. Furthermore, to test whether and
how the possible misalignment between galaxies and host halos may
affect our results for dark matter halos, we have performed  a simple
experiment in which we artifically asign a misalignment  to the
orientation of the halos, assuming a Gaussian distribution function
for the misalignment angle with a width of 35 degrees. With such
misalignment being included, the alignment statistics for the halos
become substantially weaker, thus agreeing better with the
observational results from the CMASS. This suggests that the
large-scale alignment detected in the BOSS data is physically real and
can be explained by the large-scale alignment of dark matter halos
with respect to  the matter distribution, as recently found by
\citet{Faltenbacher-Li-Wang-12} from cosmological simulations (also
see a follow-up work by \citealt{Papai-Sheth-13} who developed a
theoretical model to explain this finding).  Detailed modeling of the
observed alignment statistics should be able to provide powerful
constraints on many aspects in both galaxy formation and structure
formation theories, and would need to include a  number of effects
that are not considered in this work, including the contamination of
satellite galaxies in the CMASS sample and the too simple mass cut in
the dark matter halo sample. We will come back to this point in next
studies.

\acknowledgments

We are grateful to the anonymous referee whose comments have helped
us to significantly improve our paper.
CL acknowledges the  support of  the 100  Talents Program  of Chinese
Academy    of    Sciences    (CAS),   Shanghai    Pujiang    Programme
(no. 11PJ1411600) and the  exchange program between Max Planck Society
and  CAS.   This  work  is  sponsored  by  NSFC  (11173045,  11233005,
10878001,   11033006,  11121062)   and  the   CAS/SAFEA  International
Partnership Program for Creative Research Teams (KJCX2-YW-T23). This
work has made use of the public data from the SDSS-III.  The MultiDark
Database used in this paper and the web application providing online
access to it were constructed as part of the activities of the German
Astrophysical Virtual Observatory as result of a collaboration between
the Leibniz-Institute for Astrophysics Potsdam (AIP) and the Spanish
MultiDark Consolider Project CSD2009-00064. The Bolshoi and MultiDark
simulations were run on the NASA's Pleiades supercomputer at the NASA
Ames Research Center.


\label{lastpage}
\end{document}